\documentclass[aps,twocolumn,prb]{revtex4-2}
\setcitestyle{super,open={},close={}}

\usepackage{amsmath}
\usepackage{textcomp}
\usepackage[pdftex]{graphicx}
\usepackage{xr} 
\usepackage{float}
\usepackage[lmargin=50pt,rmargin=50pt,tmargin=50pt,bmargin=50pt]{geometry}

\usepackage[usenames]{color}

\begin{document}
	
	\title{Transient 2D IR spectroscopy and multiscale simulations reveal vibrational couplings in the Cyanobacteriochrome Slr1393-g3}
	
\author{David Buhrke$^{1,2,*}$, Yigal Lahav$^{3,4}$, Aditya Rao$^{3}$, Jeannette Ruf$^{1}$, Igor Schapiro$^{3}$ and Peter Hamm$^{1}$\\
\textit{$^1$Department of Chemistry, University of Zurich, 8057 Zurich, Switzerland\\
$^2$Institute of Biology, Humboldt University Berlin, 10115 Berlin, Germany\\
$^3$Fritz Haber Center for Molecular Dynamics, Hebrew University of Jerusalem, 9190401 Jerusalem, Israel\\
$^4$ MIGAL - Galilee Research Institute, 1101602 Kiryat Shmona, Israel\\}
$^*$corresponding author: david.buhrke@hu-berlin.de
}

\date{\today}

\begin{abstract}

 \noindent\textbf{Abstract}
Cyanobacteriochromes are bistable photoreceptor proteins with desirable photochemical properties for biotechnological applications such as optogenetics or fluorescence microscopy. Here, we investigate Slr1393-g3, a cyanobacteriochrome that reversibly photo-switches between a red-absorbing (Pr) and green-absorbing (Pg) form. We applied advanced IR spectroscopic methods to track the sequence of intermediates during the photocycle over many orders in magnitude in time. In the conversion from Pg to Pr, we have revealed a new intermediate with distinct spectroscopic features in the IR, which precedes the Pr formation by using transient IR spectroscopy. In addition, stationary and transient 2D~IR experiments measured the vibrational couplings between different groups of the chromophore and the protein in these intermediate states as well as their structural disorder. Anharmonic QM/MM calculations predict spectra in good agreement with experimental 2D~IR spectra of the initial and the final state of the photocycle. They facilitate the assignment of the IR spectra that serves as a basis for the interpretation of the spectroscopic results and suggests structural changes of the intermediates along the photocycle.
\end{abstract}

\maketitle

\section{Introduction}

\begin{figure*}[htbp]
	\centering
	\includegraphics[width=0.75\linewidth]{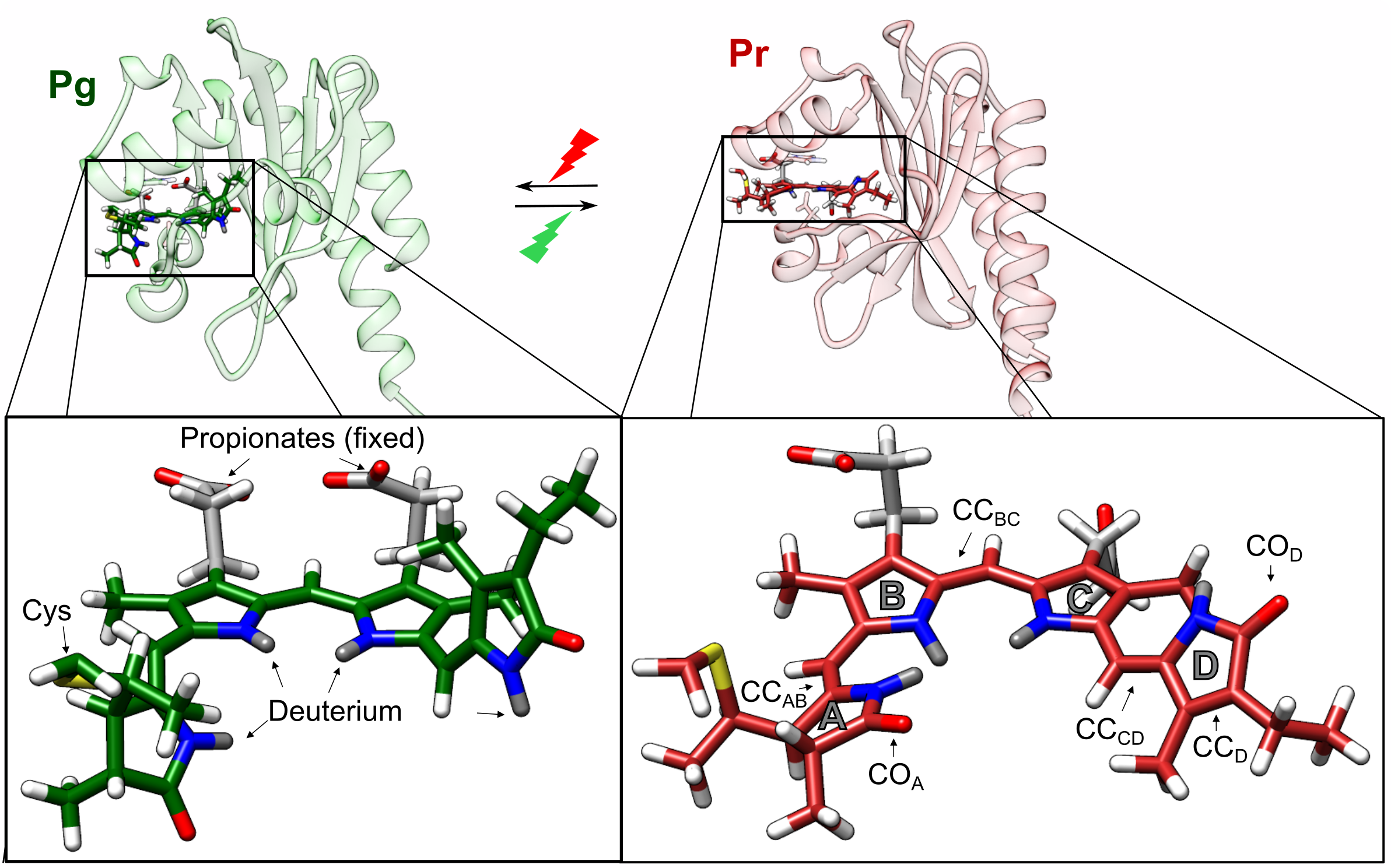}
	\caption{The QM/MM partitioning and optimized structures of Slr-g3 in the Pg state (left panel) and Pr state (right panel). The chromophore is comprised of 86 atoms (shown in licorice representation). The pyrrole hydrogens were deuterated (shown in dark gray) and positions of the propionate atoms were fixed (shown in light gray) during the calculation of vibrational frequencies. The nomenclature of the four pyrrole rings \textit{A-D} and localisation of the relevant normal modes are indicated in the right panel. 
 }
 \label{intro}
\end{figure*}

\noindent Cyanobacteriochromes (CBCRs) are bistable cyanobacterial photoreceptors\cite{Rockwell2010, Fushimi2019} with emerging applications in biotechnology, such as in super resolution microscopy\cite{Oliinyk2019} or optogenetics.\cite{Blain-Hartung2018} CBCRs typically consist of multiple photosensory modules that each incorporate an open chain tetrapyrrole chromophore such as phycocyanobilin (PCB) and are linked to cyclases, kinases or phosphodiesterases that act on cyanobacterial signaling pathways.\cite{Fushimi2019} In the present study, we investigated Slr1393-g3 (from here on denoted Slr-g3 for simplicity), a photosensory domain from a red/green histidine kinase CBCR that incorporates PCB and converts reversibly between a red-absorbing (Pr) and a green-absorbing (Pg) parent state.\cite{Chen2012, Slavov2015, Xu2014, Xu2020}
Both states are modulated by geometric changes of the PCB chromophore that are stabilized by the protein environment.
Figure \ref{intro} shows the structure of PCB inside the binding pocket of Slr-g3, which adopts a highly distorted geometry in Pg, thereby shortening the effective conjugation length compared to Pr.\cite{Xu2020, Wiebeler2018} The exact course of events that leads to these geometric changes is not fully established, hence a detailed mechanistic understanding of the photocycle reactions in red/green CBCRs is still lacking. Time-resolved spectroscopy studies performed mainly in the Larsen and Lagarias groups indicated that these proteins show large shifts in their visible absorption properties on the \textmu\text{s}-ms time scale after photoexcitation\cite{Xu2014, Jenkins2019, Kirpich2019,Kirpich2019a,Kirpich2021}, but it is up to now largely unclear how exactly PCB changes its structure as a function of time and how this process is driven by the protein environment.

Infrared (IR) spectroscopy is a powerful technique to study exactly such protein reaction dynamics, because in contrast other spectroscopic techniques like resonance Raman or visible spectroscopy it can probe the entire protein and instead of being restricted to the chromophore.\cite{Kottke2017,Lorenz-fonfria} One of the most popular observables in the mid-IR is the so-called amide I region between 1600 and 1700~cm$^{-1}$, that is dominated by signals from the peptide bond C=O stretching vibrations (called amide I modes).\cite{Barth2002} Additionally, co-factors such as PCB reveal structure-sensitive vibrational modes in the same spectral region, resulting in small IR signals superimposed on the overwhelming amide I band. In photoreceptors, these are e.g. the C=C, C=O or C=N stretching vibrations of retinal, tetrapyrrole or other co-factors that can be identified in light-dependent difference-IR experiments. The same vibrational modes can be measured in resonance Raman (RR) experiments, where they are resonance enhanced over the amide I background. Thus,  RR and IR can be used for cross-validation and provide assignments for difference-IR experiments.\cite{Buhrke2019a, Lorenz-fonfria} The amide~I region of different proteins was investigated in detail not only by linear (1D) but also two dimensional (2D-)IR spectroscopy, which offers additional information such as spectral diffusion, vibrational couplings or anharmonicities.\cite{GANIM2007} Generally, the analysis of the amide I region is tedious due to a strong coupling of the broad amide I modes, and due to spectral congestion, which makes it extremely difficult to extract local information even for small proteins and peptides.\cite{Hamm1998, Strasfeld2009, GANIM2007} Due to this obstacle, most applications of 2D~IR spectroscopy on protein samples require carefully designed experimental conditions such as isotope labeling and very small systems,\cite{Bredenbeck2003,Shim2009} or the introduction of IR labels that absorb outside the amide vibrations, typically in the so-called ``transparent window'' between 1800-2300~cm$^{-1}$.\cite{Adhikary2017,Thielges2021} Although 2D~IR spectroscopy has been thoroughly developed over the last two decades, the technique is still non-standard and requires highly specialized labs and on top of that very elaborate equipment such as femtosecond laser systems and optical parametric amplifiers. Due to a combination of these challenges, 2D~IR spectroscopy has so far been applied  only sparsely to investigate the light-induced reactions of photoreceptor proteins, e.g. on Bacteriorhodopsin\cite{Andresen2009, Hamm2021}, Photoactive Yellow Protein\cite{Schmidt-Engler2020}, a CBCR\cite{Ruf2021}, and phytochromes.\cite{Buhrke2022,Chenchiliyan2023}

Here, we tracked the IR response of Slr-g3 after excitation of Pg in the \textmu\text{s}-ms time range with transient IR spectroscopy; we will call this rather conventional method transient 1D~IR spectroscopy from now on. Further, we show that the transformation of the chromophore geometry is accompanied by changes in the coupling between certain vibrational normal modes that show up as difference cross-peaks in a stationary difference-2D~IR experiment. We then combined these two approaches and investigated the photorecation with transient 2D~IR spectroscopy in the \textmu\text{s}-ms time range.\cite{Hamm2021} Through the combination of these three related IR methods with quantum mechanics/molecular mechanics (QM/MM) simulations and anharmonic frequency calculations, we were able to develop a solid mechanistic model for the photocycle of Slr-g3.

\section{Materials and Methods}

\subsection{Protein expression and purification}

Slr-g3 was expressed in \textit{E. coli} with an \textit{in vivo} chromophore assembly protocol and purified as described previously.\cite{Buhrke2020b} For all IR experiments, the samples were dissolved in 50~mM Tris, 200~mM NaCl D$_2$O buffer (pD=7.8) and equilibrated over night to allow for complete H/D exchange of all protic hydrogens of the protein and at the four pyrrole hydrogens\cite{Buhrke2020} and had a concentration of ca. 2~mM as determined by their UV absorption at 280~nm.

\subsection{IR Spectroscopy}

All IR experiments were performed with the same experimental setup.\cite{Hamm2021} A commercial 100~kHz laser system (Amplitude) and optical parameteric amplifier (FastLite) were used to generate fs mid-IR pulses. The IR pulses were split into pump, probe and reference beams, the latter two of which were passed through the sample cuvette separated by roughly 1~mm, and then imaged onto the entrance slit of a spectrograph that was equipped with a 2x32-element MCT array detection and single-shot ADC electronics at 100~kHz.\cite{Farrell2020} In most experiments, the enhanced referencing scheme introduced in Ref.~\onlinecite{Feng2017} has been applied.

For transient 1D~IR spectroscopy, the IR pump pulses were blocked, while green actinic pulses were derived from a Nd:YAG laser (AO-S-532, CNI, Changchun, China, pulse duration ca. 40~ns) that was electronically synchronized to the 100 kHz probe light. Transient 1D~IR spectra were recorded in a vis-pump-IR-multiprobe scheme.\cite{Greetham2016,Hamm2021,Jankovic2021b,Buhrke2021} The green pump pulse energy was 17~\textmu\text{J} and the polarization was set to magic angle relative the probe beam to suppress orientational effects. The spot size of the pump beam was 180 \textmu{m} FWHM and the excitation density 670 J/m$^2$. A baseline correction was employed to account for heating effects in the sample by using the kinetic trace at 1770~cm$^{-1}$ that is largely free of vibrational signals (details are given in supplementary note 1).

For stationary 2D~IR spectroscopy, the actinic pump pulse was blocked, while a pair of IR pump pulses generated in a pulse shaper (PhaseTech) were imaged onto the sample.\cite{Middleton2010} We used 4-state phase cycling\cite{Shim2009a} to suppress scattering and strong undersampling in a rotating frame to minimize the number of laser shots (down to 168) needed to measure a complete 2D~IR spectrum. The waiting time between pump pulses and probe pulse was set to 200~fs.

Finally, for transient 2D~IR spectroscopy, both the IR-pump pulse pair as well as the actinic pump were imaged onto the sample. We used an interleaved sampling of 2D~IR and actinic pump delays, as introduced in Ref.~\onlinecite{Hamm2021}, which allowed us to generate a sequence of transient 2D~IR  spectra by proper reshuffling the data, each separated by 10~\textmu\text{s}. The transient 2D~IR  spectra were binned on a logarithmic time-scale with 10 time points per decade, rendering the signal-noise ratio at later delay times better than at earlier times since more transient 2D~IR  spectra enter a bin.

For the transient 1D~IR experiment, the repetition rate of the actinic pump laser, which determines the measurable time window, has been 15~Hz. It has been increased to 30~Hz for the transient 2D~IR  measurements, just enough to cover the complete reaction cycle, again in order to minimize the number of laser shots needed to measure a complete sequence of 2D~IR spectra (ca. 600000).

The sample was contained in a closed-cycle flow system consisting of a peristaltic pump and a sample flow cell with solenoid-driven micro valves developed for precise sample exchange synchronized to the laser system.\cite{Buhrke2021} The flow cycle contained an additional cell between the peristaltic pump and the pressure reservoir that allowed to prepare samples in the Pg state (ca. 90\%\cite{Buhrke2020b}) by illumination with a red laser diode (HL6750MG, Thorlabs) before entering the measurement cell (path length 50~\textmu\text{m}).

\subsection{QM/MM calculations}

The models used for the simulations were  based on the crystal structure of Slr-g3 and prepared as previously described.\cite{5dfx,Wiebeler2019,Wiebeler2019_2,Rao2021} Briefly, the structures of both Pr and Pg forms (PDB IDs: 5DFX and 5M82, respectively) were protonated and embedded in a TIP3P\cite{tip3p} water box, followed by MM optimization keeping the chromophore restrained to its crystal structure geometry. In this work, starting with the MM optimized structures, all solvent atoms distant more than 3.5~$\mathrm{\AA}$ of the chromophore were removed from the structure and a preliminary QM/MM optimization was performed at XTB2\cite{Grimme19}:ff14SB\cite{Simmerling15} level of theory using the ORCA\cite{ORCA} software package. The QM region contains 86 atoms (Fig. \ref{intro}), which includes the chromophore and the side chain of its binding cysteine residue. A hydrogen link atom is used to saturate the truncated bond between the C$_{\alpha}$ and C$_{\beta}$ carbons of the cysteine sidechain. A second QM/MM optimization was carried out at the B3LYP\cite{Becke92}/6-31G*\cite{Pople98}:ff14SB level of theory with Grimme's D3 dispersion correction with Becke-Johnson damping\cite{Grimme11} using the Gaussian16\cite{Gaussian16} software package. During this step, all MM atoms were kept fixed to their spatial positions, and the PCB pyrrole hydrogens were replaced by deuterium atoms. The optimized structures were subject to IR spectra calculation using the harmonic approximation. In this step the propionate groups of the chromophore were fixed. For calculating the 2D IR spectrum we followed the procedure by Mukamel et al.\cite{Mukamel2003,Mukamel2003_2}, focusing on the harmonic modes between 1600 and 1800~cm$^{-1}$. These include the two C=O stretches of rings A and D, the three C=C stretches of the methine bridges, and the C=C stretch inside ring D (right panel of Fig. \ref{intro}). These modes were subject to anharmonic frequencies, overtones and combination bands, calculation using second order perturbation theory\cite{Barone2004,Barone2005}, as implemented in the Gaussian16 package. It was validated by Geva et. al.\cite{Geva09} against full diagonalization of the vibronic Hamiltonian and was within good agreement. The final 2D IR signal was calculated by non-linear response theory\cite{hamm11} using the anharmonic frequencies and the harmonic IR intensities. The homogeneous dephasing time T$_2$ was taken as 1.0~ps. 

\begin{figure*}[htbp]
	\centering
	\includegraphics[width=0.9\linewidth]{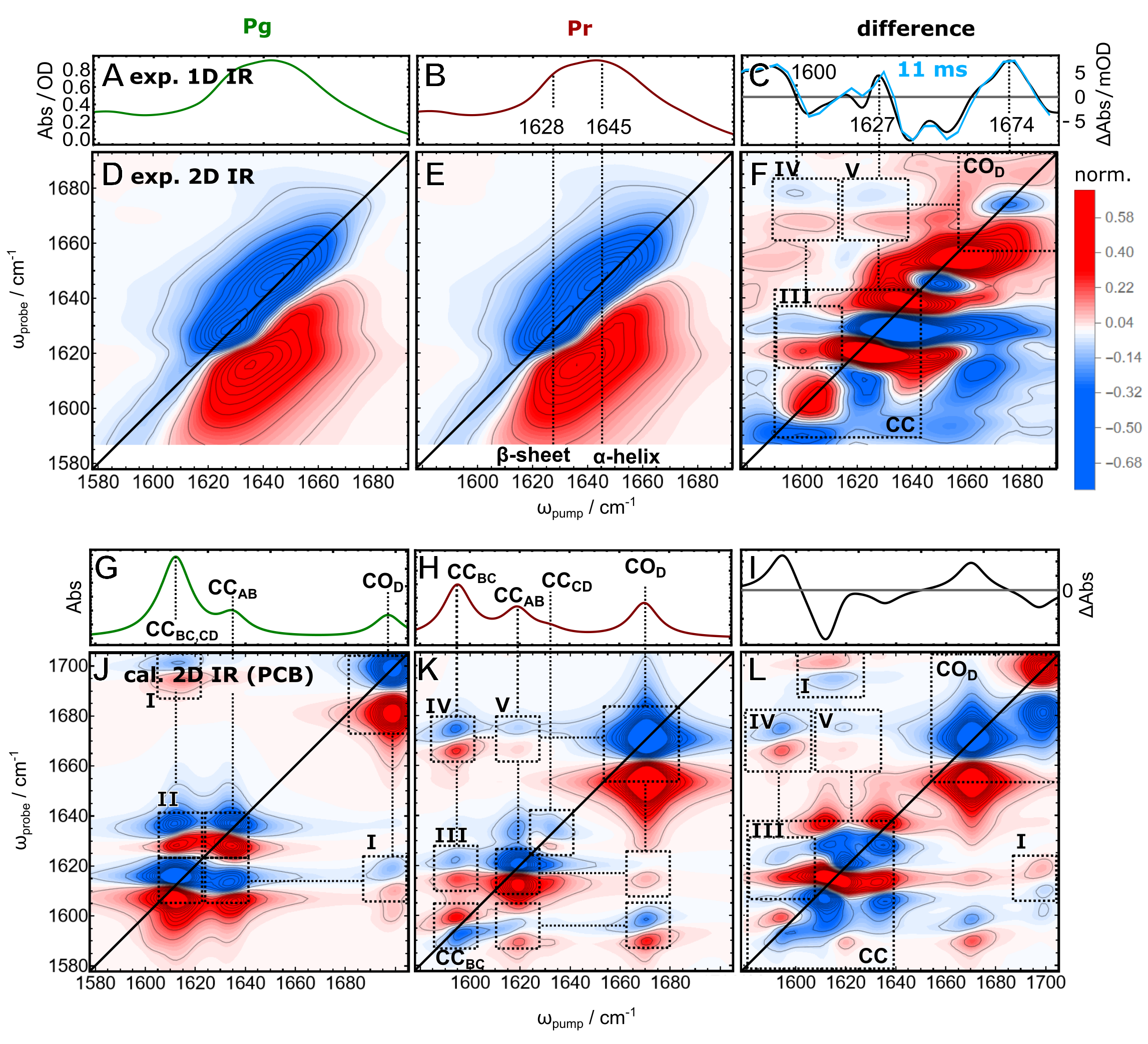}
	\caption{Stationary 1D~IR and 2D~IR spectroscopy of Slr-g3 in the amide I region. (A,B) 1D~IR spectra in the Pg and Pr states, respectively, and (C) the ``Pr-minus-Pg'' difference 1D IR spectrum. The blue line in panel (C) compares the staionary difference spectrum with the late time transient difference spectrum measured 11~ms after excitation. Panels (D)-(F) show the corresponding 2D~IR spectra, where blue colors indicate the negative signal for the 0-1 transition of the corresponding oscillators, and red colors positive signals from the anharmonically shifted 1-2-transitions. Calculated 1D~IR spectra of PCB in the Pg and Pr states, as well as the ``Pr-minus-Pg'' difference spectrum are shown in panels (G)-(I) and the corresponding 2D~IR spectra in panels (J)-(L).Note that the frequency range in panels G-L is slightly larger than in panels A-F to account for all modes relevant for the experimental data.}
	\label{2DIR}
\end{figure*}

\section{Results and Discussion}

\subsection{Stationary difference spectroscopy}

\begin{table*}
\caption{Experimental and calculated anharmonic frequencies and intensities of relevant vibrational modes in the Pg/Pr states. Vibrational modes that could not be detected experimentally are denoted n.d. (no data).}
\begin{tabular}{ |p{2.5cm}||p{3cm}|p{3.5cm}|p{4cm}|p{6cm}|  }
\hline
Mode & Exp. freq. (cm$^{-1}$) & Calc. freq. (cm$^{-1}$) & Calc. IR int. (km mol$^{-1}$) \\
 \hline
CO$_\mathrm{A}$   & n.d./1720  &  1762/1724 & 356/310\\
CO$_\mathrm{D}$   & 1700/1675  &  1697/1670 & 714/1117\\
CC$_\mathrm{AB}$  & 1637\cite{Buhrke2020}/1627  &  1635/1619 & 647/846\\
CC$_\mathrm{CD}$  & n.d./n.d.  &  1617/1632 & 203/192\\
CC$_\mathrm{BC}$  & n.d./n.d. &  1612/1595 & 2301/1637\\
CC$_\mathrm{D}$  & n.d./n.d.   &  1605/1594 & 12/24 \\
 \hline
\end{tabular}
\label{Table}
\end{table*}

The 1D IR absorption spectrum of Slr-g3 is dominated by amide I vibrations in the form of an asymmetric, broad and unstructured band in the region between 1600 and 1700~cm$^{-1}$. The spectra in the Pg and Pr states appear to be identical because the spectral changes associated with the photoconversion are too small to see them on the scale of Fig.~\ref{2DIR}A and B. The small changes associated with the photoconversion can be extracted by calculating a  ``Pr-minus-Pg'' difference 1D~IR spectrum, see Fig.~\ref{2DIR}C. Here, negative features correspond to loss of absorption due to Pg bleach, while positive bands indicate an absorption gain in Pr.

The 2D~IR spectra of Slr-g3 in the Pr and the Pg states are  indistinguishable as well in the representation of Fig~\ref{2DIR}D and E. All peaks in a 2D~IR spectrum appear as pairs with a negative (blue) lobe at higher probe frequencies and a positive (red) lobe at lower probe frequencies. Due to the quadratic dependence of the 2D~IR signal on the absorption cross section and the resulting increase in spectral resolving power, two peaks at 1628~cm$^{-1}$ and 1645~cm$^{-1}$ can be distinguished on the diagonal that only appear as slight asymmetry in the 1D~IR absorption spectra. These main amide I peaks are tentatively assigned to contributions of $\alpha$-helical and $\beta$-sheet structures of the protein, respectively.\cite{Barth2007}

Similar to the difference 1D~IR spectrum, a ``Pr-minus-Pg'' difference 2D~IR spectrum was calculated, which bears information about light-induced structural changes (Fig.~\ref{2DIR}F). The difference 2D~IR spectrum is about 10 times smaller than the 2D~IR spectra of Slr-g3 in the Pr and the Pg states (Fig. \ref{2DIR}D, E), and  is significantly richer, with many distinct diagonal and cross peaks.  All features in the difference 1D~IR spectrum have counterparts on the diagonal of the difference 2D~IR spectrum that correspond to the change in IR absorbance at the respective frequencies caused by the photoconversion of the protein. However, each diagonal peak consist of two contributions, one on the diagonal corresponding to the 0-1-transition of the corresponding vibrational mode and a second one below the diagonal with opposite sign due to the anharmonically shifted 1-2-transition (excited state absorption).\cite{hamm11} Depending on the signs of the various bands in the difference 1D~IR spectrum, the signs of the diagonal peaks vary.

\begin{figure}[t]
	\centering
	\includegraphics[width=0.75\linewidth]{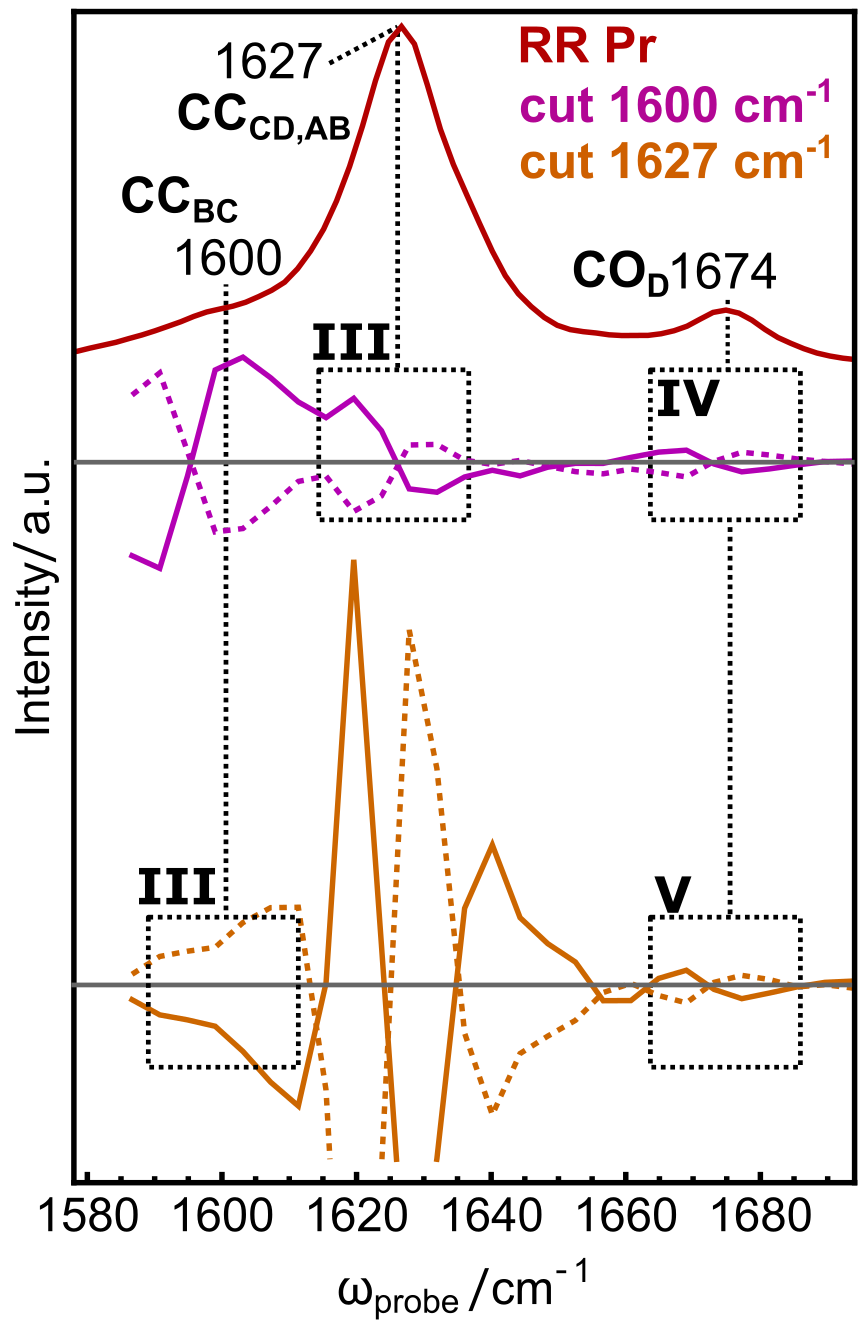}
	\caption{Vertical cuts through the experimental difference 2D~IR spectrum at 1600~cm$^{-1}$ (magenta) and 1627~cm$^{-1}$ (orange). The respective dotted lines are taken from a separate ``Pg-minus-Pr'' control experiment for reversibility. Red line: RR spectrum of Slr-g3 in the Pr state (reproduced from ref.~\onlinecite{Buhrke2020}).}
	\label{Raman}
\end{figure}

In addition, cross peaks are observed in the difference 2D~IR spectrum, labelled  \textbf{III},  \textbf{IV} and \textbf{V} in Fig. \ref{2DIR}F. The signs of all these cross peaks indicate they originate from Pr. In order to verify these cross peaks despite their small size (they are more than a factor 100 smaller than the 2D-IR spectra of Pg and Pr), the sample was converted back and forth between both states multiple times, and the inverted ``Pg-minus-Pr'' difference spectrum was also calculated. The reversibility and relative size of the cross-peaks can best be judged in vertical cuts at $\omega_{pump}$ at their maxima shown in Fig.~\ref{Raman}.  Here, the solid magenta and orange lines represent cuts through the ``Pr-minus-Pg'' spectrum while dashed lines correspond to an independently measured, reversed ``Pr-minus-Pg'' experiment.

We also compare these cuts in Fig.~\ref{Raman} with the experimental resonance Raman (RR) spectrum of Slr-g3 in the Pr state that we reproduced from Ref.~\onlinecite{Buhrke2020}. The frequencies of the cross- and diagonal peaks coincide with the main peaks of the RR spectrum, confirming that these signals originate from the conjugated C=C system. In Fig.~\ref{2DIR}F, the difference cross-peaks are clearly visible only on one side of the main diagonal with  $\omega_{probe}>\omega_{pump}$, while they are masked by the stronger excited state absorption contributions of the diagonal signals on the other side.  Nevertheless, in Fig.~\ref{Raman}, a superimposed cross-peak can be identified as a kink at 1600~cm$^{-1}$ in the vertical cut at 1627~cm$^{-1}$ (orange line, labeled \textbf{III}).

We anticipate that most features in the difference spectra originate from the pronounced geometric changes of the PCB chromophore, while some must also originate from the changes of protein structure that have impact on the amide I spectrum.\cite{Buhrke2020b} These signals generally overlap with each other. To distinguish these contributions, and to facilitate an assignment, we performed anharmonic QM/MM calculations where the protein environment was fixed and only vibrational modes from the chromophore were considered. Six vibrational modes of PCB were found in the investigated frequency window between 1580 and 1750~cm$^{-1}$, the two, mostly localized C=O stretching modes of the ring A and D carbonyls as well as C=C stretching modes of the conjugated system. The latter are  dominated by stretching motions of the three methine bridges (A-B, B-C and C-D), and are denoted accordingly as CC$_\mathrm{XY}$ in the following. One mode inside ring D (CC$_\mathrm{D}$, see Table~\ref{Table}) was identified as well, but its intensity is very weak, hence this mode will not be discussed here. The calculated atomic displacement vectors of all identified modes in Pg and Pr are given in Fig.S1.  

The calculated frequencies and  IR intensities are listed in Table~\ref{Table} for both states Pg/Pr, the calculated anharmonicities are listed in Tables S1 and S2, and displacement vectors are given in figure S1. Fig.~\ref{2DIR}G and H plot the resulting 1D~IR spectra, which are dominated by 3 peaks (i.e., CC$_\mathrm{BC}$ and CC$_\mathrm{AB}$ as well as CO$_\mathrm{D}$). CC$_\mathrm{CD}$ is very weak and visible only in Pr as a shoulder, while CO$_\mathrm{A}$ is outside the spectral window of Fig.~\ref{2DIR}. The finding that CO$_\mathrm{A}$ absorbs at a higher frequency than CO$_\mathrm{D}$ is in good agreement with the literature on PCB-binding proteins and originates mainly from the different level of saturation in the pyrrole rings A and D.\cite{Foerstendorf2001, Ruf2021}

With the exception of  CC$_\mathrm{CD}$, all modes reveal a frequency downshift upon the Pg-to-Pr transition.

Based on the QM/MM results, and in combination of the Raman spectrum of Pr shown in Fig.~\ref{Raman},\cite{Buhrke2020} we can assign the experimental peaks of the Pr state, see Table~\ref{Table} (we will return to the Pg state later). This assignment is based on the energy ordering and relative intensities of the  various modes. It is interesting to note that among the bridging modes,  CC$_\mathrm{BC}$ has the strongest IR intensities while CC$_\mathrm{AB}$ has the stronger Raman intensity (Fig.~\ref{Raman}). In accordance with the QM/MM results, this implies that both modes are delocalized among each other to a certain degree, with one being more of a symmetric character and the other of a asymmetric character. Based on the results of the calculations, we conclude that CC$_\mathrm{CD}$ is too weak to be identified experimentally.

Each of the calculated resonances in the 1D~IR spectra shows up with the characteristic doublet on the the diagonal of the corresponding 2D~IR spectrum of Pg (Fig.~\ref{2DIR}J). In addition, strong cross peaks \textbf{I} (between CO$_\mathrm{D}$ and the CC modes) and \textbf{II} (between CC$_\mathrm{BC}$ and CC$_\mathrm{AB}$) reveal that the chromophore modes are coupled among each other.

The 2D~IR spectrum of Pr is better resolved (Fig.~\ref{2DIR}K), since the C=C resonances are spread over a wider frequency range. In particular, a cross peak \textbf{III}  can be identified, connecting the strongly coupled CC$_\mathrm{BC}$ and CC$_\mathrm{AB}$ modes. Furthermore,  CO$_\mathrm{D}$ now produces two distinct cross peaks with the CC$_\mathrm{BC}$ (\textbf{IV}) and CC$_\mathrm{AB}$ (\textbf{V}). These spectral changes give rise to a specific pattern in the calculated difference spectrum (Fig. \ref{2DIR}L), reproducing the main features of the experimental difference spectrum surprisingly well (Fig. \ref{2DIR}F). This is in particular true for the cross peaks \textbf{III}, \textbf{IV} and \textbf{V} that connect the major signals from the C=C region among each other and with  CO$_\mathrm{D}$ in the Pr state. The  cross peak \textbf{I} with opposite sign, is not visible in the experimental spectrum, either because it is too weak or outside the measured frequency window.

\begin{figure}[t]
	\centering
	\includegraphics[width=\linewidth]{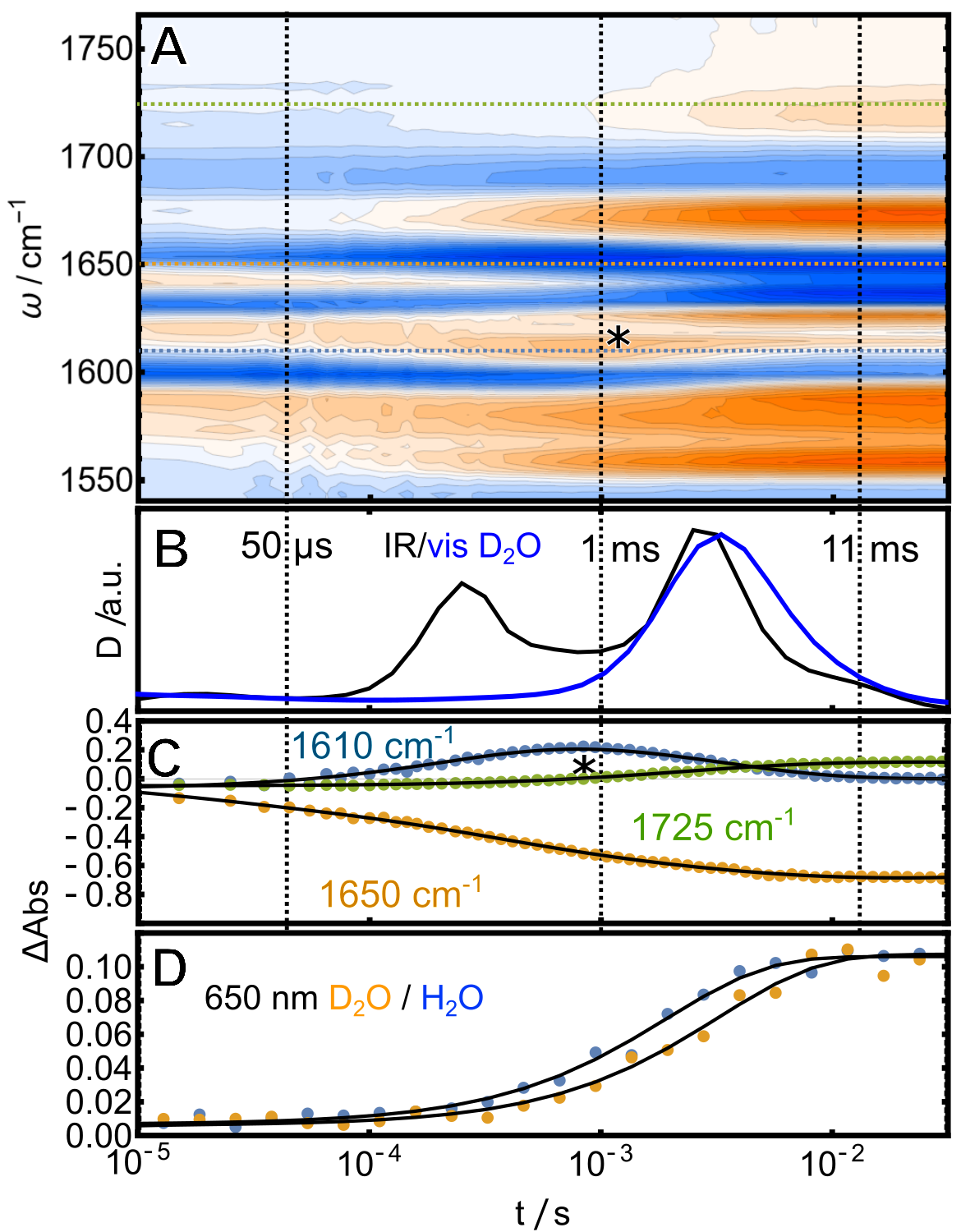}
	\caption{Transient 1D~IR spectroscopy of the Pg $\rightarrow$ Pr reaction from 10~\textmu\text{s} to 30~ms. (A) Data represented as a contour plot; blue colors indicate negative bands (bleaches) and red colors positive bands. (B) Dynamical content of the IR data compared to the transient visible data recorded in D$_2$O (blue trace). (C) Three representative kinetic traces with lifetime analysis fits (black lines). (D) Representative kinetic trace in the visible 650~nm recorded in H$_2$O (blue) and D$_2$O buffer (orange).}
	\label{T1DIR}
\end{figure}

\subsection{Transient 1D~IR spectroscopy}

The stationary 1D~IR and 2D~IR difference spectra of Figs.~\ref{2DIR}C and F report on changes in the geometric and electronic structure between the two stable states of the protein. The time-resolved counterparts of these spectroscopies can provide the same information for transient intermediates during the course of the photocycle. In a recent publication,\cite{Buhrke2020b} we presented transient 1D~IR data of Slr-g3 in the ps to \textmu\text{s} time range obtained with a pump-probe setup based on two electronically synchronized femtosecond laser systems.\cite{Bredenbeck2004} Here, we add new data in the \textmu\text{s}-ms regime acquired with a conceptually different vis-pump-IR-multiprobe spectrometer, operating at repetition rates of 15~Hz (vis pump) and 100~kHz (IR probe, see Materials and Methods).\cite{Hamm2021,Jankovic2021b,Buhrke2021}
The new data reveal two further reaction steps that can be identified directly by visual inspection of the contour plot of Fig.~\ref{T1DIR}A, or in the three selected traces shown in Fig.~\ref{T1DIR}C. Here, a local maximum at 1610~cm$^{-1}$ and 1~ms clearly indicates the formation and decay of an intermediate state with a distinct spectral signature (marked with an asterisk). Time constants for these processes were obtained by lifetime analysis with a maximum entropy method,\cite{Kumar2001,Lorenz-Fonfria:06} similar to the analysis that was applied for the first part of the data in our recent paper (technical details are given in supplementary note 2).\cite{Buhrke2020b} The resulting dynamical content $D$ is plotted in Fig.~\ref{T1DIR}B, which reveals two clear maxima at 250~\textmu\text{s} and 2.5~ms associated with the time constants of these processes.  The properly scaled late time (i.e., 11~ms) transient 1D~IR spectrum is overlaid in Fig.~\ref{2DIR}C (blue line) with the stationary 1D~IR difference spectrum (black line). The near-perfect agreement of both spectra indicates that the photoreaction is completed after that time and Pr is formed.

The micro-to-millisecond kinetics of Slr-g3 were up to now only studied by transient visible spectroscopy.\cite{Xu2014} We reproduced the visible experiment in the region from 350 to 700~nm in H$_2$O and D$_2$O buffer (fig. S2), and overlaid the dynamical content of the D$_2$O data with the IR data in figure \ref{T1DIR} B. Here, we observed only one clear maximum at a time that is in good agreement with the second process observed in the IR. This indicates that the intermediate state preceding Pr cannot be distinguished in the visible experiment and only the last process has significant impact on the visible spectrum. Xu et. al reported two time constants (340~\textmu{s} and 1.03~ms) from a global fit of visible data on the same time scales that are surprisingly in good agreement with the two time constants we observe in the IR.\cite{Xu2014} However, these authors did not resolve an intermediate state with a distinct spectral signature, instead all their visible spectra in the micro- to millisecond regime intersect at one isosbestic point. We also analyzed our visible data with a global fit (fig. S2), and found two H/D sensitive time constants at 1.3~\textmu{s} and 1.9~ms but could not reproduce a time constant with several hundred \textmu{s}. 

\begin{figure*}[t]
	\centering
	\includegraphics[width=\linewidth]{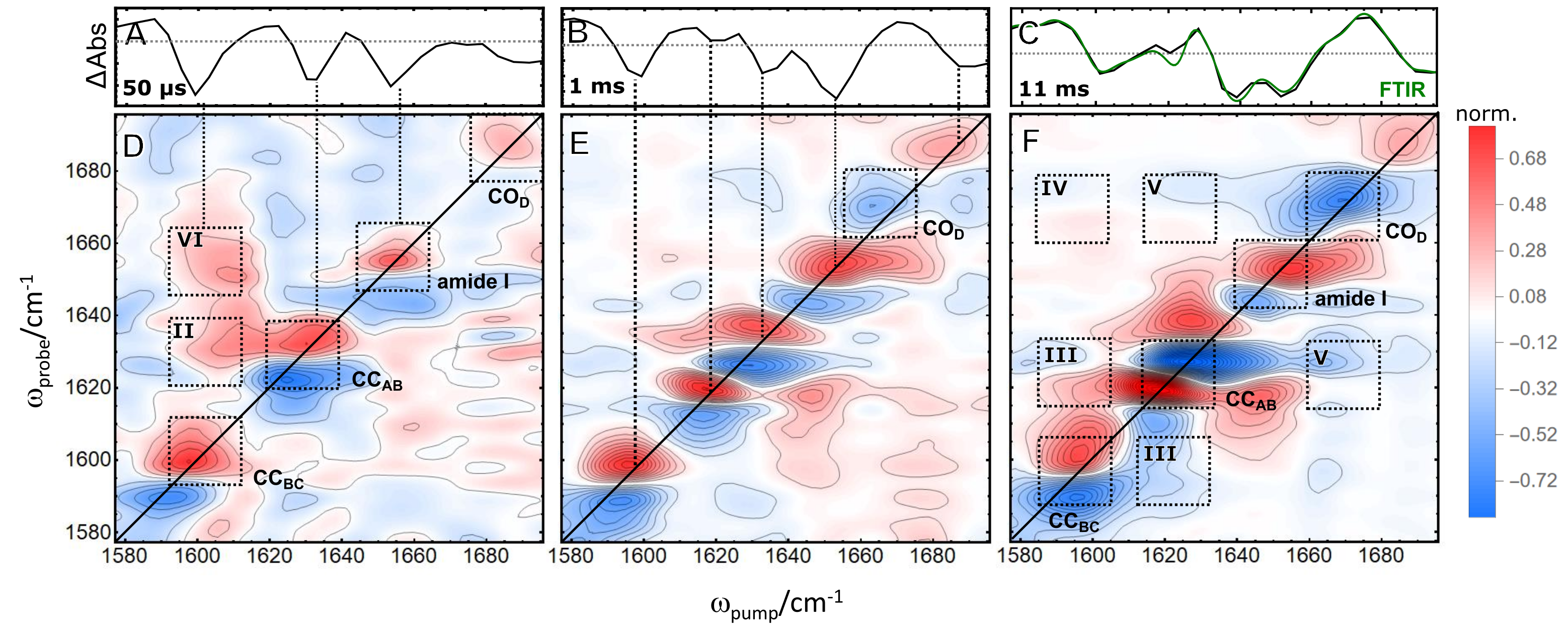}
	\caption{Transient 1D~IR (top row) and transient 2D~IR spectra at time-points (50~\textmu\text{s}, 1~ms, and 11~ms) at which the population of certain intermediates is maximized, see Fig.~\ref{T1DIR}C. Transient 1D~IR spectra are taken from the data set of Fig.~\ref{T1DIR}A.}
	\label{FigTransient2DIR}
\end{figure*}

Fig.~\ref{FigTransient2DIR}A-C shows transient 1D~IR spectra at time points 50~\textmu\text{s}, 1~ms, and 11~ms, at which the populations of the various intermediates are maximized. In the 50~\textmu\text{s} intermediate (Fig.~\ref{FigTransient2DIR}A), the bleach bands are stronger by a factor 2x-5x than their positive counterparts. Based on the results of the QM/MM calculations, we assign the two lower-frequency bleaches to the two bridging modes CC$_\mathrm{BC}$ and CC$_\mathrm{AB}$, while CC$_\mathrm{CD}$ is expected to be too weak to contribute significantly. The CO$_\mathrm{D}$ band at 1695~cm$^{-1}$ completely lacks a positive signal in the 1D spectrum. Based on the QM/MM results, we also conclude that the negative band at around 1655~cm$^{-1}$ cannot originate from the chromophore, hence it must originate from the protein. Indeed, in our previous study,\cite{Buhrke2020b} we observed that the kinetic process leading to that band on a 40~ns timescales is only observed in the IR transient spectra, and not in the electronic spectra of the chromophore, which lead us to conclude that it reflects a response of the protein, probably the unfolding of a short helical segment.

The predominant bleach character of the 50~\textmu\text{s} spectrum is likely due to disorder in the PCB chromophore, leading to broadened positive bands with weak intensity. Additionally, the kinetic isotope effect in our H/D sensitive flash photolysis data hints that (de-)protonation events likely play a role in the micro- to millisecond kinetics (figs. \ref{T1DIR}D and S2), and therefore this intermediate state could also involve deprotonation of PCB. However, deprotonation and structural disorder must not be logically linked with one another and might just coincide incidentally. Transient de- and re-protonation processes involving the tetrapyrrole chromophores were also found in the related red/green CBCR AnPixJ\cite{Escobar2013, Song2015} and are also typical for later intermediates in the photocycle of phytochromes.\cite{VanThor2001,Borucki2005} According to NMR and RR experiments, the PCB chromophore of red/green CBCRs can deprotonate either at the ring B or C pyrrole nitrogen, resulting in a tautomeric species where only one proton is shared between these two pyrrole rings.\cite{Song2015, Altmayer2022}

The next intermediate is maximized at 1~ms (Fig.~\ref{FigTransient2DIR}B), and shows two distinct differences compared to the 50~\textmu{s} species. First, the region around 1620~cm$^{-1}$ now shows two dispersive features instead of the one dominant bleach (dotted vertical lines highlight the maxima). There are two possible reasons for this band shape. Either these signals originate from the frequency shift of two different normal modes of the chromophore, which we find unlikely since the IR intensity of the CC$_\mathrm{CD}$ mode in that region is predicted to be small. 
Alternatively, it reflects a change in protein structure on this time scale, which produces an amide I difference signal. Since the 250~\textmu{s} process is not detected in the visible spectra, or, is very small in the best case,\cite{Xu2014} we exclude the possibility of changes in the C=C stretching modes and assign this feature to the protein. The second notable difference is that a positive counterpart to the CO$_\mathrm{D}$ bleach appears at the frequency position of the Pr state. Together with the first assignment, this provides evidence that the change in protein structure establishes the electrostatic environment around CO$_\mathrm{D}$, which persists in the Pr state. 

The last reaction step occurs with ca. 2.5~ms, in good agreement with our D$_2$O data in the visible region. The formation of Pr in the early milliseconds is preceded by an orange-absorbing intermediate state (Fig. S2),\cite{Xu2014} and a cryo-trapping RR study found that ring D is already in a Pr-like configuration while ring A still retains an out-of-plane tilt similar to Pg.\cite{Buhrke2020} Such a configuration has a conjugated system with an size that is in-between that of Pg and Pr, hence an electronic absorption band between green and red. The transient 1D~IR results of Fig. \ref{T1DIR} validate the hypothesis that the last reaction step is associated with a movement of ring A, because here it can be directly observed that the positive CO$_\mathrm{A}$ at $>$1700~cm$^{-1}$ is only established in the last reaction step (green trace in Fig. \ref{T1DIR}C). Since this reaction step also reveals an kinetic isotope effect (Fig.~\ref{T1DIR}D), and PCB is protonated in Pr,\cite{Buhrke2020} the last step is likely connected to the reprotonation of the chromophore.

\subsection{Transient 2D~IR spectroscopy}

Figs.~\ref{FigTransient2DIR}D-F show a series of transient 2D~IR spectra at the same time points as the corresponding 1D~IR spectra. Just like for transient 1D~IR spectroscopy (Fig.~\ref{2DIR}C, black vs. blue line), the late time transient 2D~IR spectrum at 11~ms is in essence the same as the stationary 2D~IR difference spectrum of Fig.~\ref{2DIR}F. With respect to the diagonal peaks, the 1~ms transient 2D~IR  spectrum is the same as the final 11~ms transient 2D~IR  spectrum. However, the 1~ms transient 2D~IR  spectrum has in essence no cross-peak features; these appear only in the final reaction step. In addition, only inverted diagonal peaks are observed in the 1~ms transient 2D~IR  spectrum, i.e., the  fundamental on the diagonal colored in red paired with an anharmonically shifted blue peak below (dotted lines). Each of the diagonal peaks is linked to a negative peak in the transient 1D~IR spectrum, originating  from the depleted Pg state.  The corresponding positive bands in the transient 1D~IR data have no counterpart on the diagonal of the transient 2D~IR spectra with a blue/red pattern (with the one exception of the CO$_\mathrm{D}$ mode). As discussed in Ref.~\onlinecite{bre03c}, in principle four peaks are expected on the diagonal for each mode in  a transient 2D~IR spectrum, the two inner ones typically merging if the frequency shift of the mode is smaller than its bandwidth. However, as also discussed in Ref.~\onlinecite{bre03c}, the  contribution from the transient intermediate is very strongly suppressed if it is structurally more disordered than the reactant state. The capability of 2D~IR spectroscopy to disentangle inhomogeneous from homogeneous broadening, combined with strong cancellation effects between those four peaks in the double-difference spectrum that constitutes a transient 2D~IR spectrum (i.e., the difference between 0-1 and 1-2 transition and the difference between intermediate and reactant state), is responsible for the suppression of signals from the intermediate.\onlinecite{bre03c} In that case, only two peaks are observed from the reactant state, with a color pattern (signs) that is the same as that observed here. We therefore conclude that the 1~ms intermediate is structurally more disordered than the Pg form.

The same applies to the 50~\textmu\text{s} transient 2D~IR spectrum, which is even simpler this regard with only four diagonal peaks with inverted signs. 
An inverted cross peak \textbf{II} can  be identified between CC$_\mathrm{BC}$ and CC$_\mathrm{AB}$. That is, as the bands on the diagonal bleach, the corresponding cross peaks bleaches as well. The calculations predict the strongest coupling between the adjacent  CC$_\mathrm{AB}$ and CC$_\mathrm{BC}$ modes (cross peak \textbf{II} in Fig. \ref{2DIR}J), in agreement with the present assignment of the  50~\textmu\text{s} transient spectra. Cross peak \textbf{VI} indicates the coupling between the chromophore and the protein (Fig.~\ref{FigTransient2DIR}A).

\section{Conclusion}

In transient IR spectroscopy of photoactive proteins, signals that originate from the chromophore are often dominant, since its electronic and geometric structure changes significantly more than that of the embedding protein.\cite{Gerwert1990, Ritter2015, Kottke2017, Lorenz-fonfria} Notable exceptions are phytochromes, LOV domains, and orange carotenoid proteins, where a large-scale re- or unfolding of entire secondary structure elements occurs.\cite{Stojkovic2014, Konold2016,Konold2019} The x-ray structures of Slr-g3 shown in Fig.~\ref{intro} reveal that the biggest changes in this protein concern rings A and D of the PCB chromophore. While all four rings are almost coplanar in the Pr state, rings A and D are tilted out of the plane in Pg. These geometries immediately explain the color change between the two states, i.e., their electronic properties: When rings A and D become co-planar with rings B and C, the three sp2-hybridized atoms that are part of ring A and five atoms of ring D become part of the conjugated system, thereby shifting its electronic absorption bands towards the red.\cite{Wiebeler2018} Additionally, some differences in secondary structure have also been identified by x-ray crystallography, such as e.g. the unfolding of a small segment of the so-called a3-helix in the Pr state.\cite{Xu2020}

The kinetic analysis of our time-resolved data indicates two processes in the \textmu\text{s} to ms time range, or in other words, two meta-stable intermediate states that are maximized around 50~\textmu\text{s} and 1~ms, respectively, before Pr is formed. Both of these meta-stable configurations are covered by the transient 2D~IR experiment, and thus the connectivity of the dominant modes allows unprecedented insight into the chromophore configuration at this stage of the photocycle.

A combination of three advanced IR methodologies allowed us to track the structural changes of the PCB chromophore inside the protein matrix of the red/green CBCR Slr-g3. First, transient 1D~IR spectroscopy was used to establish the time scales of various photocycle events. The ability of IR spectroscopy to sense structural changes that do not impact the electronic transitions probed by UV/Vis spectroscopy revealed an additional intermediate
preceding the formation of Pr, which has not been observed before. Here it is important to note that the discussed vibrational modes overlap with the strongly overwhelming amide I modes from the protein backbone. Nevertheless, the vibratonal modes of the chromophore can be isolated by difference spectroscopy and validated by the calculations. On the other hand, this approach identifies contributions from the protein backbone that are linked to protein function. Comparison of transient IR with transient VIS spectroscopy allows to disentangle kinetic steps that are associated mostly with the protein vs those that involve the chromophore; only the latter are detectable in the vis. Finally, transient 2D~IR difference spectra of various intermediate species were obtained, that allowed us to gain insights into the coupling patterns in the various intermediates, as well as their structural disorder. The combination of these four techniques allowed us to draw consistent conclusions about the reaction mechanism that Slr-g3 undergoes after excitation. For example, reaction steps could be determined during which a Pr-like electrostatic environment is established around ring D and during which ring A rotates into plane to form the Pr state.

The work also demonstrates that a close-to-quantitative agreement between 2D-IR spectra and anharmonic QM/MM calculations  can be obtained for the initial and final forms, Pg and Pr, respectively. The existing X-ray structures of these two forms served as a basis of the QM/MM simulations. Here, we also make suggestions for the structural changes of the two transient intermediates preceding the formation of final Pr state, based on a combination of  transient 1D and 2D spectroscopy. Time-resolved X-ray scattering experiments are nowadays within reach, using upcoming X-ray free-electron laser facilities,\cite{Standfuss2019} which could provide the fundament of rigorous QM/MM calculations also of those intermediates, to test our structural hypotheses. Since Slr-g3 (as well as other CBCRs) are emerging tools in biotechnological applications, such as optogenetics and fluorescence microscopy,\cite{Oliinyk2019,Blain-Hartung2018} it is highly desirable to understand the molecular mechanism of its photoreaction in real time and at full atomistic detail.\\

\noindent \textbf{Acknowledgements:} This work was supported by the Swiss National Science Foundation (SNF) through Grant No. 200020B 188694/1. DB acknowledges a Liebig-Scholarship by the funds of the German chemical industry (Fonds der chemischen Industrie, FCI). I.S. acknowledges support from the Israel Ministry of Science and Technology (Grant 3-16311). I.S. thanks the DFG Collaborative Research Center 1078, project C6 for support. This project emerged during the preparation of the grant application for SNF Sinergia 213507.\\

\noindent \textbf{Supporting information:} The supporting information for this study contains two tables with the calculated anharmonicities of the overtones and combination bands, a figure of the respective displacement vectors and a figure with further information on the flash photolysis data. The SI also contains two supplementary notes with details on the baseline correction of the transient IR data and the lifetime analysis.\\

\noindent \textbf{Data availability statement:} The data that support the findings of this study are available at 10.5281/zenodo.8113340.\\

\bibliography{refs}

\begin{thebibliography}{10}

\bibitem{Rockwell2010}
Rockwell NC, Lagarias JC.
\newblock {A brief history of phytochromes}.
\newblock ChemPhysChem. 2010;11(6):1172-80.

\bibitem{Fushimi2019}
Fushimi K, Narikawa R.
\newblock {Cyanobacteriochromes: photoreceptors covering the entire
  UV-to-visible spectrum}.
\newblock Current Opinion in Structural Biology. 2019;57:39-46.
\newblock Available from: \url{https://doi.org/10.1016/j.sbi.2019.01.018}.

\bibitem{Oliinyk2019}
Oliinyk OS, Shemetov AA, Pletnev S, Shcherbakova DM, Verkhusha VV.
\newblock {Smallest near-infrared fluorescent protein evolved from
  cyanobacteriochrome as versatile tag for spectral multiplexing}.
\newblock Nature Communications. 2019;10(1):1-13.
\newblock Available from: \url{http://dx.doi.org/10.1038/s41467-018-08050-8}.

\bibitem{Blain-Hartung2018}
Blain-Hartung M, Rockwell NC, Moreno MV, Martin SS, Gan F, Bryant DA, et~al.
\newblock {Cyanobacteriochrome-based photoswitchable adenylyl cyclases (cPACs)
  for broad spectrum light regulation of cAMP levels in cells}.
\newblock Journal of Biological Chemistry. 2018;293(22):8473-83.

\bibitem{Chen2012}
Chen Y, Zhang J, Luo J, Tu JM, Zeng XL, Xie J, et~al.
\newblock {Photophysical diversity of two novel cyanobacteriochromes with
  phycocyanobilin chromophores: Photochemistry and dark reversion kinetics}.
\newblock FEBS Journal. 2012;279(1):40-54.

\bibitem{Slavov2015}
Slavov C, Xu X, Zhao KH, G{\"{a}}rtner W, Wachtveitl J.
\newblock {Detailed insight into the ultrafast photoconversion of the
  cyanobacteriochrome Slr1393 from Synechocystis sp.}
\newblock Biochim Biophys Acta - Bioenerg. 2015;1847(10):1335-44.

\bibitem{Xu2014}
Xu XL, Gutt A, Mechelke J, Raffelberg S, Tang K, Miao D, et~al.
\newblock {Combined mutagenesis and kinetics characterization of the
  bilin-binding GAF domain of the protein Slr1393 from the cyanobacterium
  synechocystis PCC6803}.
\newblock ChemBioChem. 2014;15(8):1190-9.

\bibitem{Xu2020}
Xu X, Port A, Wiebeler C, Zhao KH, Schapiro I, G{\"{a}}rtner W.
\newblock {Structural elements regulating the photochromicity in a
  cyanobacteriochrome}.
\newblock Proceedings of the National Academy of Sciences of the United States
  of America. 2020;117(5):2432-40.

\bibitem{Wiebeler2018}
Wiebeler C, {Rao} AG, G{\"{a}}rtner W, Schapiro I.
\newblock {The Effective Conjugation Length is Responsible for the Red/Green
  Spectral Tuning in the Cyanobacteriochrome Slr1393g3}.
\newblock Angewandte Chemie International Edition. 2018;58(7):1934-8.
\newblock Available from: \url{http://doi.wiley.com/10.1002/anie.201810266}.

\bibitem{Jenkins2019}
Jenkins AJ, Gottlieb SM, Chang CW, Hayer RJ, Martin SS, Lagarias JC, et~al.
\newblock {Conservation and diversity in the secondary forward photodynamics of
  red/green cyanobacteriochromes}.
\newblock Photochemical and Photobiological Sciences. 2019;18(10):2539-52.

\bibitem{Kirpich2019}
Kirpich JS, Gottlieb SM, Chang CW, Kim PW, Martin SS, Lagarias JC, et~al.
\newblock {Forward Photodynamics of the Noncanonical Red/Green NpR3784
  Cyanobacteriochrome from Nostoc punctiforme}.
\newblock Biochemistry. 2019;58(18):2297-306.

\bibitem{Kirpich2019a}
Kirpich JS, Gottlieb SM, Chang CW, Kim PW, Martin SS, Lagarias JC, et~al.
\newblock {Reverse Photodynamics of the Noncanonical Red/Green NpR3784
  Cyanobacteriochrome from Nostoc punctiforme}.
\newblock Biochemistry. 2019;58(18):2307-17.

\bibitem{Kirpich2021}
Kirpich JS, Chang CW, Franse J, Yu Q, Escobar FV, Jenkins AJ, et~al.
\newblock {Comparison of the Forward and Reverse Photocycle Dynamics of Two
  Highly Similar Canonical Red/Green Cyanobacteriochromes Reveals Unexpected
  Differences}.
\newblock Biochemistry. 2021;60:274-88.

\bibitem{Kottke2017}
Kottke T, L{\'{o}}renz-Fonfr{\'{i}}a VA, Heberle J.
\newblock {The grateful infrared: Sequential protein structural changes
  resolved by infrared difference spectroscopy}.
\newblock Journal of Physical Chemistry A. 2017;121(2):335-50.

\bibitem{Lorenz-fonfria}
Lorenz-Fonfria VA.
\newblock {Infrared Difference Spectroscopy of Proteins: From Bands to Bonds}.
\newblock Chem Rev. 2020;120(7):3466–3576.

\bibitem{Barth2002}
Barth A, Zscherp C.
\newblock {What vibrations tell about proteins}.
\newblock Q Rev Biophys. 2002;35(4):370.

\bibitem{Buhrke2019a}
Buhrke D, Hildebrandt P.
\newblock {Probing Structure and Reaction Dynamics of Proteins Using Time-
  Resolved Resonance Raman Spectroscopy}.
\newblock Chemical Reviews. 2019:3577-630.

\bibitem{GANIM2007}
Ganim Z, Chung HS, Smith AW, Deflores LP, Jones KC, Tokmakoff A.
\newblock {Amide I Two-Dimensional Infrared Spectroscopy of Proteins}.
\newblock Accounts of Chemical Research. 2007;41(3):432-44.

\bibitem{Hamm1998}
Hamm P, Lim M, Hochstrasser RM.
\newblock {Structure of the Amide I Band of Peptides Measured by Femtosecond
  Nonlinear-Infrared Spectroscopy}.
\newblock The Journal of Physical Chemistry B. 1998;102(31):6123-38.
\newblock Available from: \url{http://pubs.acs.org/doi/abs/10.1021/jp9813286}.

\bibitem{Strasfeld2009}
Strasfeld DB, Ling YL, Gupta R, Raleigh DP, Zanni MT.
\newblock {Strategies for Extracting Structural Information from 2D IR
  Spectroscopy of Amyloid : Application to Islet Amyloid}.
\newblock J Phys Chem B. 2009;113(47):15679-91.

\bibitem{Bredenbeck2003}
Bredenbeck J, Hamm P.
\newblock {Peptide structure determination by two-dimensional infrared
  spectroscopy in the presence of homogeneous and inhomogeneous broadening}.
\newblock Journal of Chemical Physics. 2003;119(3):1569-78.

\bibitem{Shim2009}
Shim SH, Gupta R, Ling YL, Strasfeld DB, Raleigh DP, Zanni MT.
\newblock {Two-dimensional IR spectroscopy and isotope labeling defines the
  pathway of amyloid formation with residue-specific resolution}.
\newblock Proceedings of the National Academy of Sciences of the United States
  of America. 2009;106(16):6614-9.

\bibitem{Adhikary2017}
Adhikary R, Zimmermann J, Romesberg FE.
\newblock {Transparent Window Vibrational Probes for the Characterization of
  Proteins with High Structural and Temporal Resolution}.
\newblock Chemical Reviews. 2017;117(3):1927-69.

\bibitem{Thielges2021}
Thielges MC.
\newblock {Transparent window 2D IR spectroscopy of proteins}.
\newblock J Chem Phys. 2021;155:40903.
\newblock Available from: \url{https://doi.org/10.1063/5.0052628}.

\bibitem{Andresen2009}
Andresen ER, Hamm P.
\newblock {Site-Specific Difference 2D-IR Spectroscopy of Bacteriorhodopsin}.
\newblock J Phys Chem B. 2009;113(18):6520-7.

\bibitem{Hamm2021}
Hamm P.
\newblock {Transient 2D IR Spectroscopy from Micro- to Milliseconds}.
\newblock J Chem Phys. 2021;154:104201.
\newblock Available from: \url{https://doi.org/10.1063/5.0045294}.

\bibitem{Schmidt-Engler2020}
Schmidt-Engler JM, Blankenburg L, Zangl R, Hoffmann J, Morgner N, Bredenbeck J.
\newblock {Local dynamics of the photo-switchable protein PYP in ground and
  signalling state probed by 2D-IR spectroscopy of-SCN labels †}.
\newblock Phys Chem Chem Phys. 2020;22:22963.

\bibitem{Ruf2021}
Ruf J, Hamm P, Buhrke D.
\newblock {Needles in a Haystack: H-bonding in an Optogenetic Protein observed
  with Isotope Labeling and 2D-IR Spectroscopy}.
\newblock Physical Chemistry Chemical Physics. 2021;23:10267  10273.

\bibitem{Buhrke2022}
Buhrke D, Michael N, Hamm P.
\newblock {Vibrational couplings between protein and cofactor in bacterial
  phytochrome Agp1 revealed by 2D-IR spectroscopy}.
\newblock Proc Natl Acad Sci USA. 2022;119(31):e2206400119.

\bibitem{Chenchiliyan2023}
Chenchiliyan M, K{\"{u}}bel J, Ooi SA, Salvadori G, Mennucci B, Westenhoff S,
  et~al.
\newblock {Ground-state heterogeneity and vibrational energy redistribution in
  bacterial phytochrome observed with femtosecond 2D}.
\newblock J Chem Phys. 2023;158(085103):1-11.
\newblock Available from: \url{https://doi.org/10.1063/5.0135268}.

\bibitem{Buhrke2020b}
Buhrke D, Oppelt KT, Heckmeier PJ, Fern{\'{a}}ndez-Ter{\'{a}}n R, Hamm P.
\newblock {Nanosecond protein dynamics in a red/green cyanobacteriochrome
  revealed by transient IR spectroscopy}.
\newblock The Journal of Chemical Physics. 2020;153(24):245101.
\newblock Available from: \url{https://doi.org/10.1063/5.0033107}.

\bibitem{Buhrke2020}
Buhrke D, Battocchio G, Wilkening S, Blain-Hartung M, Baumann T, Schmitt FJ,
  et~al.
\newblock {Red, Orange, Green: Light- And Temperature-Dependent Color Tuning in
  a Cyanobacteriochrome}.
\newblock Biochemistry. 2020;59:509-19.

\bibitem{Farrell2020}
Farrell KM, Ostrander JS, Jones AC, Yakami BR, Dicke SS, Middleton CT, et~al.
\newblock {Shot-to-shot 2D IR spectroscopy at 100 kHz using a Yb laser and
  custom-designed electronics}.
\newblock Opt Express. 2020;28(22):33584.

\bibitem{Feng2017}
Feng Y, Vinogradov I, Ge NH.
\newblock {General noise suppression scheme with reference detection in
  heterodyne nonlinear spectroscopy}.
\newblock Optics Express. 2017;25(21):26262.

\bibitem{Greetham2016}
Greetham GM, Donaldson PM, Nation C, Sazanovich IV, Clark IP, Shaw DJ, et~al.
\newblock {A 100 kHz time-resolved multiple-probe femtosecond to second
  infrared absorption spectrometer}.
\newblock Applied Spectroscopy. 2016;70(4):645-53.

\bibitem{Jankovic2021b}
Jankovic B, Ruf J, Zanobini C, Bozovic O, Buhrke D, Hamm P.
\newblock {Sequence of Events During Peptide Unbinding from RNase S: A Complete
  Experimental Description}.
\newblock J Phys Chem Lett. 2021;12:5201-7.

\bibitem{Buhrke2021}
Buhrke D, Ruf J, Heckmeier P, Hamm P.
\newblock {A stop-flow sample delivery system for transient spectroscopy}.
\newblock Review of Scientific Instruments. 2021;92(12):123001.
\newblock Available from: \url{https://doi.org/10.1063/5.0068227}.

\bibitem{Middleton2010}
Middleton CT, Woys AM, Mukherjee SS, Zanni MT.
\newblock {Residue-specific structural kinetics of proteins through the union
  of isotope labeling, mid-IR pulse shaping, and coherent 2D IR spectroscopy}.
\newblock Methods. 2010;52(1):12-22.
\newblock Available from: \url{http://dx.doi.org/10.1016/j.ymeth.2010.05.002}.

\bibitem{Shim2009a}
Shim SH, Zanni MT.
\newblock {How to turn your pump-probe instrument into a multidimensional
  spectrometer: 2D IR and Vis spectroscopies via pulse shaping}.
\newblock Physical Chemistry Chemical Physics. 2009;11(5):748-61.

\bibitem{5dfx}
Xu X, Höppner A, Wiebeler C, Zhao KH, Schapiro I, Gärtner W.
\newblock Structural elements regulating the photochromicity in a
  cyanobacteriochrome.
\newblock Proceedings of the National Academy of Sciences. 2020;117(5):2432-40.
\newblock Available from:
  \url{https://www.pnas.org/doi/abs/10.1073/pnas.1910208117}.

\bibitem{Wiebeler2019}
Wiebeler C, Schapiro I.
\newblock {QM/MM Benchmarking of Cyanobacteriochrome Slr1393g3 Absorption
  Spectra}.
\newblock Molecules. 2019;24(9):1720.
\newblock Available from: \url{https://www.mdpi.com/1420-3049/24/9/1720}.

\bibitem{Wiebeler2019_2}
Wiebeler C, Rao AG, Gärtner W, Schapiro I.
\newblock The Effective Conjugation Length Is Responsible for the Red/Green
  Spectral Tuning in the Cyanobacteriochrome Slr1393g3.
\newblock Angewandte Chemie International Edition. 2019;58(7):1934-8.

\bibitem{Rao2021}
Rao AG, Wiebeler C, Sen S, Cerutti DS, Schapiro I.
\newblock Histidine protonation controls structural heterogeneity in the
  cyanobacteriochrome AnPixJg2.
\newblock Phys Chem Chem Phys. 2021;23:7359-67.
\newblock Available from: \url{http://dx.doi.org/10.1039/D0CP05314G}.

\bibitem{tip3p}
Jorgensen WL, Chandrasekhar J, Madura JD, Impey RW, Klein ML.
\newblock Comparison of simple potential functions for simulating liquid water.
\newblock The Journal of Chemical Physics. 1983;79(2):926-35.
\newblock Available from: \url{https://doi.org/10.1063/1.445869}.

\bibitem{Grimme19}
Bannwarth C, Ehlert S, Grimme S.
\newblock GFN2-xTB—An Accurate and Broadly Parametrized Self-Consistent
  Tight-Binding Quantum Chemical Method with Multipole Electrostatics and
  Density-Dependent Dispersion Contributions.
\newblock Journal of Chemical Theory and Computation. 2019;15(3):1652-71.
\newblock PMID: 30741547.
\newblock Available from: \url{https://doi.org/10.1021/acs.jctc.8b01176}.

\bibitem{Simmerling15}
Maier JA, Martinez C, Kasavajhala K, Wickstrom L, Hauser KE, Simmerling C.
\newblock ff14SB: Improving the Accuracy of Protein Side Chain and Backbone
  Parameters from ff99SB.
\newblock Journal of Chemical Theory and Computation. 2015;11(8):3696-713.
\newblock PMID: 26574453.
\newblock Available from: \url{https://doi.org/10.1021/acs.jctc.5b00255}.

\bibitem{ORCA}
Neese F, Wennmohs F, Becker U, Riplinger C.
\newblock The ORCA quantum chemistry program package.
\newblock The Journal of Chemical Physics. 2020;152(22):224108.
\newblock Available from: \url{https://doi.org/10.1063/5.0004608}.

\bibitem{Becke92}
Becke AD.
\newblock Density‐functional thermochemistry. I. The effect of the
  exchange‐only gradient correction.
\newblock The Journal of Chemical Physics. 1992;96(3):2155-60.
\newblock Available from: \url{https://doi.org/10.1063/1.462066}.

\bibitem{Pople98}
Rassolov VA, Pople JA, Ratner MA, Windus TL.
\newblock 6-31G* basis set for atoms K through Zn.
\newblock The Journal of Chemical Physics. 1998;109(4):1223-9.
\newblock Available from: \url{https://doi.org/10.1063/1.476673}.

\bibitem{Grimme11}
Grimme S, Ehrlich S, Goerigk L.
\newblock Effect of the damping function in dispersion corrected density
  functional theory.
\newblock Journal of Computational Chemistry. 2011;32(7):1456-65.
\newblock Available from:
  \url{https://onlinelibrary.wiley.com/doi/abs/10.1002/jcc.21759}.

\bibitem{Gaussian16}
Frisch MJ, Trucks GW, Schlegel HB, Scuseria GE, Robb MA, Cheeseman JR, et~al..
  Gaussian˜16 {R}evision {C}.01; 2016.
\newblock Gaussian Inc. Wallingford CT.

\bibitem{Mukamel2003}
Hayashi T, Mukamel S.
\newblock Multidimensional Infrared Signatures of Intramolecular Hydrogen
  Bonding in Malonaldehyde.
\newblock The Journal of Physical Chemistry A. 2003;107(43):9113-31.
\newblock Available from: \url{https://doi.org/10.1021/jp030626m}.

\bibitem{Mukamel2003_2}
Moran AM, Dreyer J, Mukamel S.
\newblock Ab initio simulation of the two-dimensional vibrational spectrum of
  dicarbonylacetylacetonato rhodium(I).
\newblock The Journal of Chemical Physics. 2003;118(3):1347-55.
\newblock Available from: \url{https://doi.org/10.1063/1.1528605}.

\bibitem{Barone2004}
Barone V.
\newblock Vibrational zero-point energies and thermodynamic functions beyond
  the harmonic approximation.
\newblock The Journal of Chemical Physics. 2004;120(7):3059-65.
\newblock Available from: \url{https://doi.org/10.1063/1.1637580}.

\bibitem{Barone2005}
Barone V.
\newblock Anharmonic vibrational properties by a fully automated second-order
  perturbative approach.
\newblock The Journal of Chemical Physics. 2005;122(1):014108.
\newblock Available from: \url{https://doi.org/10.1063/1.1824881}.

\bibitem{Geva09}
Baiz CR, McRobbie PL, Preketes NK, Kubarych KJ, Geva E.
\newblock Two-Dimensional Infrared Spectroscopy of Dimanganese Decacarbonyl and
  Its Photoproducts: An Ab Initio Study.
\newblock The Journal of Physical Chemistry A. 2009;113(35):9617-23.

\bibitem{hamm11}
Hamm P, Zanni MT.
\newblock {Concepts and Methods of 2D Infrared Spectroscopy}.
\newblock Cambridge: Cambridge University Press; 2011.

\bibitem{Barth2007}
Barth A.
\newblock {Infrared spectroscopy of proteins}.
\newblock Biochimica et Biophysica Acta - Bioenergetics. 2007;1767(9):1073-101.

\bibitem{Foerstendorf2001}
Foerstendorf H, Benda C, G{\"{a}}rtner W, Storf M, Scheer H, Siebert F.
\newblock {FTIR studies of phytochrome photoreactions reveal the C=O bands of
  the chromophore: Consequences for its protonation states, conformation, and
  protein interaction}.
\newblock Biochemistry. 2001;40(49):14952-9.

\bibitem{Bredenbeck2004}
Bredenbeck J, Helbing J, Hamm P.
\newblock {Continuous scanning from picoseconds to microseconds in time
  resolved linear and nonlinear spectroscopy}.
\newblock Review of Scientific Instruments. 2004;75(11):4462-6.

\bibitem{Kumar2001}
Kumar ATN, Zhu L, Christian JF, Demidov AA, Champion PM.
\newblock {On the rate distribution analysis of kinetic data using the maximum
  entropy method: Applications to myoglobin relaxation on the nanosecond and
  femtosecond timescales}.
\newblock Journal of Physical Chemistry B. 2001;105(32):7847-56.

\bibitem{Lorenz-Fonfria:06}
L\'{o}renz-Fonfr\'{i}a VA, Kandori H.
\newblock Transformation of Time-Resolved Spectra to Lifetime-Resolved Spectra
  by Maximum Entropy Inversion of the Laplace Transform.
\newblock Appl Spectrosc. 2006 Apr;60(4):407-17.
\newblock Available from: \url{http://as.osa.org/abstract.cfm?URI=as-60-4-407}.

\bibitem{Escobar2013}
{Vel{\'{a}}zquez Escobar} F, Utesch T, Narikawa R, Ikeuchi M, Mroginski MA,
  G{\"{a}}rtner W, et~al.
\newblock {Photoconversion mechanism of the second GAF domain of
  cyanobacteriochrome AnPixJ and the cofactor structure of its green-absorbing
  state}.
\newblock Biochemistry. 2013;52(29):4871-80.

\bibitem{Song2015}
Song C, {Vel{\'{a}}zquez Escobar} F, Xu XL, Narikawa R, Ikeuchi M, Siebert F,
  et~al.
\newblock {A Red/Green Cyanobacteriochrome Sustains Its Color Despite a Change
  in the Bilin Chromophors Protonation State}.
\newblock Biochemistry. 2015;54(38):5839-48.

\bibitem{VanThor2001}
{van Thor} JJ, Borucki B, Crielaard W, Otto H, Lamparter T, Hughes J, et~al.
\newblock {Light-induced proton release and proton uptake reactions in the
  cyanobacterial phytochrome Cph1}.
\newblock Biochemistry. 2001;40(38):11460-71.

\bibitem{Borucki2005}
Borucki B, {von Stetten} D, Seibeck S, Lamparter T, Michael N, Mroginski MA,
  et~al.
\newblock {Light-induced proton release of phytochrome is coupled to the
  transient deprotonation of the tetrapyrrole chromophore}.
\newblock J Biol Chem. 2005;280(40):34358-64.

\bibitem{Altmayer2022}
Altmayer S, K{\"{o}}hler L, Bielytskyi P, G{\"{a}}rtner W, Matysik J, Wiebeler
  C, et~al.
\newblock {Light- and pH-dependent structural changes in cyanobacteriochrome
  AnPixJg2}.
\newblock Photochemical and Photobiological Sciences. 2022;21(4):447-69.
\newblock Available from: \url{https://doi.org/10.1007/s43630-022-00204-4}.

\bibitem{bre03c}
Bredenbeck J, Helbing J, Renner C, Behrendt R, Moroder L, Wachtveitl J, et~al.
\newblock {Transient 2D-IR Spectroscopy: Snapshots of the Nonequilibrium
  Ensemble during the Picosecond Conformational Transition of a Small Peptide}.
\newblock J Phys Chem B. 2003;107:8654-60.

\bibitem{Gerwert1990}
Gerwert K, Souvignier G, Hess B.
\newblock {Simultaneous monitoring of light-induced changes in protein
  side-group protonation, chromophore isomerization, and backbone motion of
  bacteriorhodopsin by time-resolved Fourier-transform infrared spectroscopy}.
\newblock Proceedings of the National Academy of Sciences of the United States
  of America. 1990;87(24):9774-8.

\bibitem{Ritter2015}
Ritter E, Puskar L, Bartl FJ, Aziz EF, Hegemann P, Schade U.
\newblock {Time-resolved infrared spectroscopic techniques as applied to
  channelrhodopsin}.
\newblock Frontiers in Molecular Biosciences. 2015;2(JUL):1-7.

\bibitem{Stojkovic2014}
Stojkovi{\'{c}} EA, Toh KC, Alexandre MTA, Baclayon M, Moffat K, Kennis JTM.
\newblock {FTIR spectroscopy revealing light-dependent refolding of the
  conserved tongue region of bacteriophytochrome}.
\newblock J Phys Chem Lett. 2014;5(15):2512-5.

\bibitem{Konold2016}
Konold PE, Mathes T, Weienborn J, Groot ML, Hegemann P, Kennis JTM.
\newblock {Unfolding of the C-Terminal J$\alpha$ Helix in the LOV2
  Photoreceptor Domain Observed by Time-Resolved Vibrational Spectroscopy}.
\newblock Journal of Physical Chemistry Letters. 2016;7(17):3472-6.

\bibitem{Konold2019}
Konold PE, {Van Stokkum} IHM, Muzzopappa F, Wilson A, Groot ML, Kirilovsky D,
  et~al.
\newblock {Photoactivation Mechanism, Timing of Protein Secondary Structure
  Dynamics and Carotenoid Translocation in the Orange Carotenoid Protein}.
\newblock Journal of the American Chemical Society. 2019;141(1):520-30.

\bibitem{Standfuss2019}
Standfuss J.
\newblock {Membrane protein dynamics studied by X-ray lasers – or why only
  time will tell}.
\newblock Curr Opin Struct Biol. 2019;57:63-71.

\end{thebibliography}
\bibliographystyle{vancouver}

\end{document}


\title{Supporting Information for: \\
 Transient 2D IR spectroscopy and multiscale simulations reveal vibrational couplings in the Cyanobacteriochrome Slr1393-g3}
	
\author{David Buhrke$^{1,2,*}$, Yigal Lahav$^{3,4}$, Aditya Rao$^{3}$, Jeannette Ruf$^{1}$, Igor Schapiro$^{3}$ and Peter Hamm$^{1}$\\
\textit{$^1$Department of Chemistry, University of Zurich, Zurich, Switzerland\\
$^2$current adress: Institute of Biology, Humboldt University Berlin, Germany\\
$^3$Fritz Haber Center for Molecular Dynamics Research Institute of Chemistry, The Hebrew University of Jerusalem, Israel\\
$^4$ MIGAL - Galilee Research Institute, Kiryat Shmona, Israel\\}
$^*$corresponding author: david.buhrke@hu-berlin.de
}

\maketitle

\begin{figure}[htbp]
	\centering
	\includegraphics[width=\linewidth]{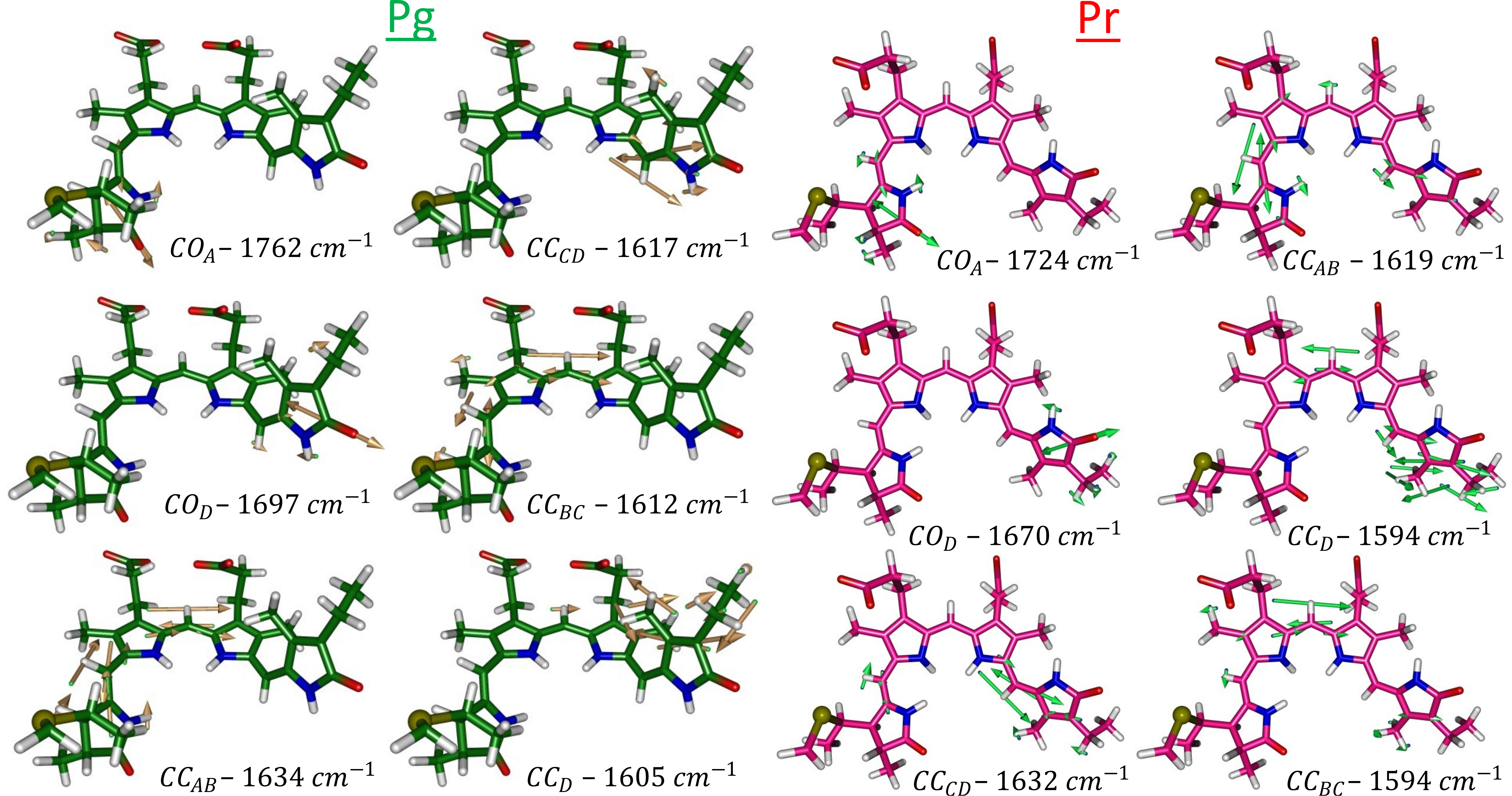}
	\caption{Displacement vectors of the calculated vibrational modes in the  Pg state (left, green) and the Pr state (right, pink).}
 \label{intro}
\end{figure}

\begin{table}

\caption{Calculated anharmonicity of the overtones (diagonal, defined as the difference between overtone and fundamental transition frequency, $\omega_{1,2}-\omega_{0,1}$) and combination bands (off-diagonal, defined as the difference between combination mode and fundamental transition frequency, $\omega_{01,11}-\omega_{00,10}$), in cm$^{-1}$ of the relevant vibrational modes in the Pg state.}

\begin{tabular}{ |p{2cm}|p{2cm}|p{2cm}|p{2cm}|p{2cm}|p{2cm}|p{2cm}|  }
\hline
  & CC$_\mathrm{D}$  & CC$_\mathrm{BC}$ & CC$_\mathrm{CD}$ & CC$_\mathrm{AB}$ & CO$_\mathrm{D}$  &  CO$_\mathrm{A}$ \\
 \hline
CC$_\mathrm{D}$  & -12.04 & -1.28 & -1.94 & -0.27 & -1.67 & 0.02 \\
\hline
CC$_\mathrm{BC}$ &  & -0.85 & -0.17 & -4.78 & -0.35 & -0.64 \\
\hline
CC$_\mathrm{CD}$ &  &  & -11.01 & -0.12 & -1.95 & 0.00 \\
\hline
CC$_\mathrm{AB}$ &  &  &  & -4.50 & -0.19 & -1.00 \\
\hline
CO$_\mathrm{D}$  &  &  &  &  & -17.13 & 0.03 \\
\hline
CO$_\mathrm{A}$  &  &  &  &  &  & -17.23 \\
  \hline
\end{tabular}
\label{TablePg}
\end{table}

\begin{table}

\caption{Calculated anharmonicity of the overtones (diagonal, defined as the difference between overtone and fundamental transition frequency, $\omega_{1,2}-\omega_{0,1}$) and combination bands (off-diagonal, defined as the difference between combination mode and fundamental transition frequency, $\omega_{01,11}-\omega_{00,10}$), in cm$^{-1}$ of the relevant vibrational modes in the Pr state.}

\begin{tabular}{ |p{2cm}|p{2cm}|p{2cm}|p{2cm}|p{2cm}|p{2cm}|p{2cm}|  }
\hline
  & CC$_\mathrm{BC}$  & CC$_\mathrm{D}$ & CC$_\mathrm{AB}$ &CC$_\mathrm{CD}$ & CO$_\mathrm{D}$  &  CO$_\mathrm{A}$ \\
 \hline
CC$_\mathrm{BC}$  & 0.46 & -2.14 & -0.56 & -0.63 & -0.57  & -0.37 \\
\hline
CC$_\mathrm{D}$  &  & -7.56 & -1.25 & -3.95 & -1.42 & 0.00 \\
\hline
CC$_\mathrm{AB}$ &  &  & -4.98 & -2.46 & -0.48 & -2.16 \\
\hline
CC$_\mathrm{CD}$ &  &  &  & -6.08 & -0.66 & -0.36 \\
\hline
CO$_\mathrm{D}$  &  &  &  &  & -16.84 & 0.06 \\
\hline
CO$_\mathrm{A}$  &  &  &  &  &  & -15.67 \\
  \hline
\end{tabular}
\label{TablePr}
\end{table}

\newpage

\begin{figure}[htbp]
	\centering
	\includegraphics[width=0.9\linewidth]{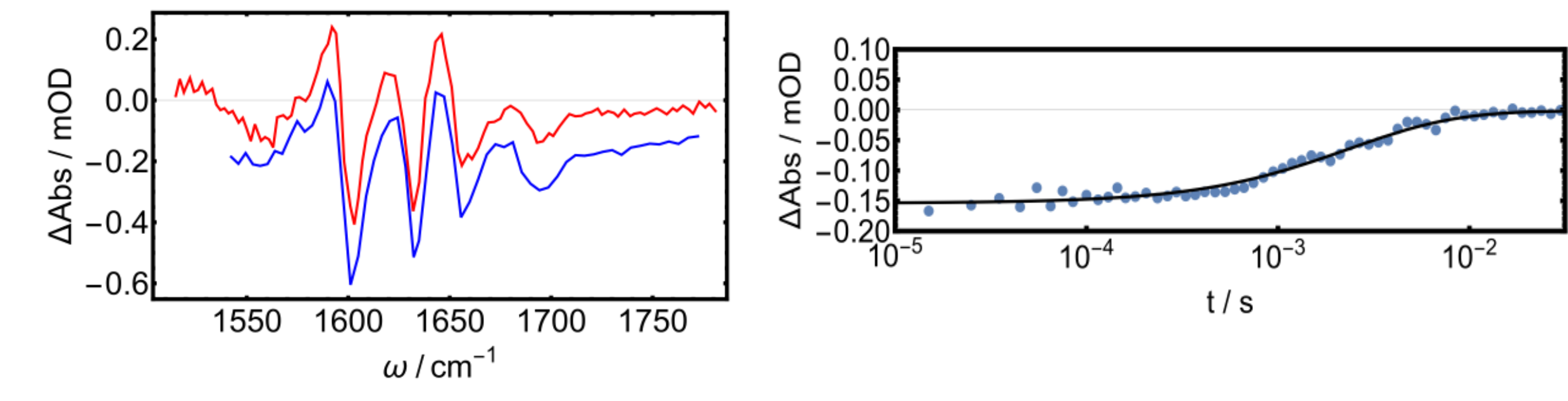}
	\caption{Left panel: pump-probe spectrum at 42~\textmu s (red line) and uncorrected vis pump / IR multi-probe spectrum at the same time (blue line). Right panel: kinetic trace of the uncorrected vis pump / IR multi-probe data at 1770~cm$^{-1}$. }
 \label{baseline}
\end{figure}

\noindent \textbf{Supplementary note 1: Baseline correction of transient IR data.} The early time points of the raw transient IR data collected with the vis-pump / IR multi-probe overlap in time with data that we reported earlier with electronically synchronized pump-probe setup and should in principle be identical. However, when the transient spectra at 42~\textmu s are overlaid, a negative offset is observed in in the raw pump / multi-probe data over the entire spectral region (fig. \ref{baseline}, left panel). We attributed this background to heating due to the strong pump pulse. We accounted for this baseline by simply subtracting the value in the signal-free region at 1770~cm$^{-1}$ from each spectrum (fig. \ref{baseline}, left panel).

\begin{figure}[htbp]
	\centering
	\includegraphics[width=0.8\linewidth]{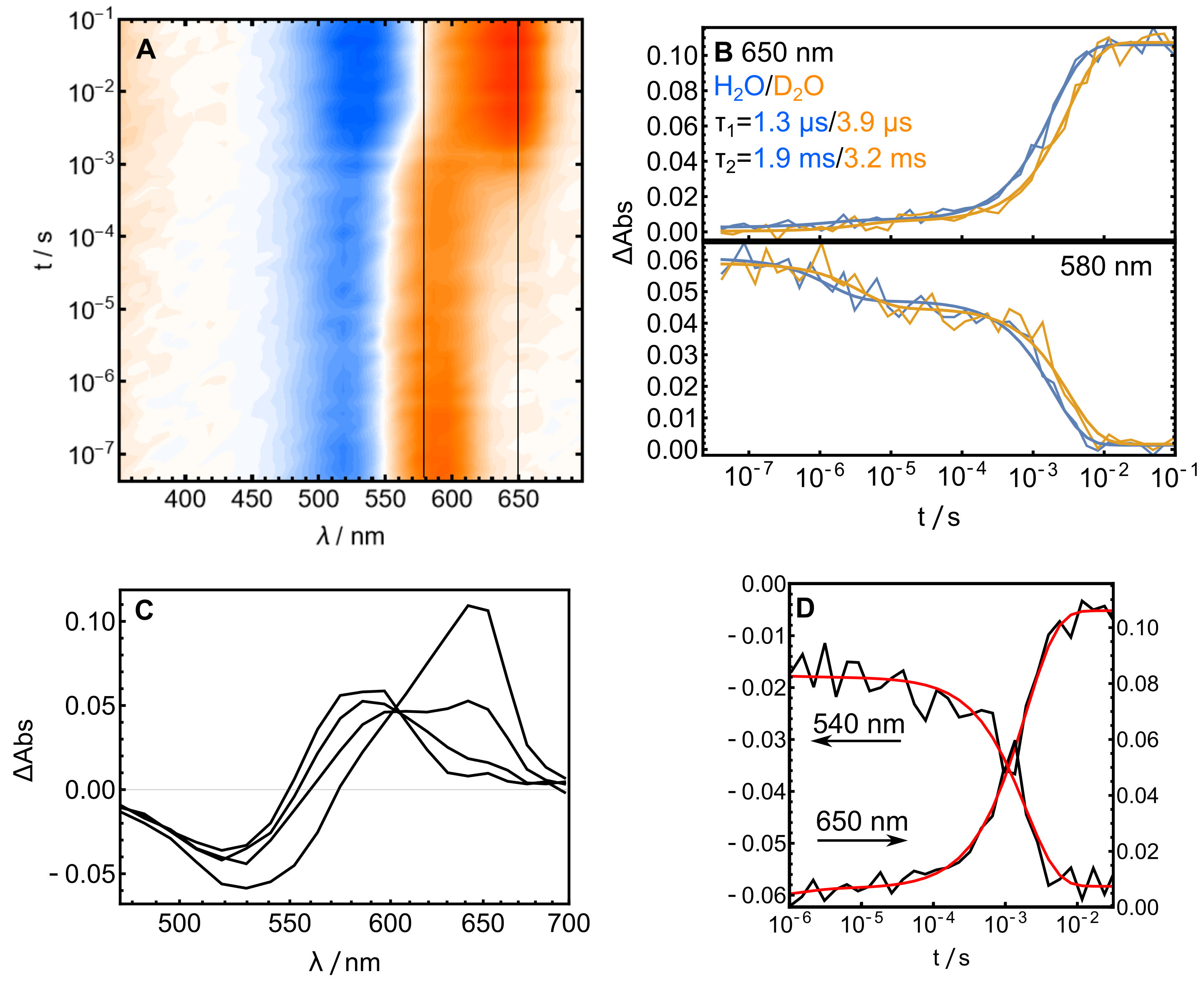}
	\caption{Flash photolysis of Slr-g3 in H$_2$O and D$_2$O. A: Data in form of a contour plot. B kinetic traces at 650 and 580 nm, showing the kinetic isotope effect on the two transitions in the early micro- and milisecond time scales. C and D: Data plotted in the same representation as in the ref. \onlinecite{Xu2014} (Fig. 7) for to ease the direct comparison with the literature.  Flash photolysis experiments were carried out with a modified LKS.60 flash-photolysis system (Applied Photophysics Ltd., Leatherhead, UK). The exciting laser pulse was tuned to 530~nm using a optical parametric oscillator (MagicPrism, OPOTEK, Carlsbad, CA, USA), pumped with the third harmonic of a Nd:YAG laser (BrilliantB, Quantel, Les Ulis, France, 10~ns pulse length).  A 150~W Xenon lamp (Osram, München, Germany) was used in conjunction with a fast shutter (UniBlitz, Rochester, NY, USA) to monitor changes in absorption.  Transient spectra were recorded using an Andor iStar ICCD camera (DH734; Andor Technology Ltd, Belfast, Ireland) with 1024 pixels. Spectra were recorded at 46 different time points between 10~ns and 100~ms with custom software written in Visual C++. After each measured time point, the sample was illuminated for 1.5~s with an home-built LED setup (680~nm) to prepare the Pg state.}
 \label{intro}
\end{figure}

\noindent \textbf{Supplementary note 2: Lifetime analysis with maximum entropy regularisation.} The T1DIR and vis data sets were analyzed with lifetime analysis. The fundamental idea behind lifetime analysis is the same as in a conventional global fit and assumes that the data can be described by interconverting discrete states with time-invariant spectra. That is, the data matrix $d(\omega_i,t_j)$ can be written as superposition of the $n$ different components:

\begin{equation}
d(\omega_i,t_j) = \sum_{k=1}^{n}C_k(t_j)A_k(\omega_i),
\label{separability}
\end{equation}

where $C_k(t_j)$ is the concentration profile of component $k$ as a function of time $t_j$, and $A_k(\omega_i)$ its spectrum at probe frequency $\omega_i$. The data were fit to multiexponential functions\cite{Hobson1998,Kumar2001,Lorenz-Fonfria:06}:

\begin{equation}
f(\omega_i,t_j)=a_0(\omega_i)-\sum_{k=1}^n a(\omega_i,\tau_k)e^{-t_j/{\tau_k}},
\label{LDA}
\end{equation}

where the index $k$ refers to a kinetic component with time constant $\tau_k$. We will abbreviate $a_{i,k}\equiv a(\omega_i,\tau_k)$.

10 time constants $\tau_k$ per decade were fixed and distributed equidistantly on a logarithmic scale, while only the amplitudes $a_{i,k}$were the free fitting parameters for each kinetic trace. A penalty function that maximizes the generalized absolute Shannon–Jones entropy $s_i$ of the amplitudes $a_{i,k}$ for each frequency $i$ was introduced to regularize the fit and avoid overfitting \cite{Lorenz-Fonfria:06,Lorenz-Fonfria2007}:

\begin{equation}
\begin{aligned}
s_i=\sum_{k}\Bigg(& \sqrt{a_{i,k}^2+4m_i^2}-a_{i,k} \mathrm{ln}\frac{\sqrt{a_{i,k}^2+4m_i^2}+a_{i,k}}{2m_i}-2m_i\Bigg)
\end{aligned}
\label{Entropy}
\end{equation}

Here, $m_i$ are the so-called \textit{a priori} solutions, which are a measure for the overall amplitude of the data at $\omega_i$ \cite{Lorenz-Fonfria:06,Lorenz-Fonfria2007}. The entropy is subtracted from the root mean square deviation $\chi^2$ of the fit, weighted by a regularisation parameter $\lambda$:

\begin{equation}
E_i=\frac{\chi^2_i}{\lambda} - s_i,
\label{error}
\end{equation}

with

\begin{equation}
\chi^2=\sum_{i,j}\left(d(\omega_i,t_j)-f(\omega_i,t_j)\right)^2.
\label{chisquare}
\end{equation}

The metric $E_i$ is minimized with respect to the amplitudes $a_{i,k}$. The sum over the squared amplitudes at all probe frequencies at one time $\tau_k$ is termed the ``dynamical content'' $D(\tau_k)$\cite{Stock2018}:

\begin{equation}
D(\tau_k) =\sqrt{\sum_{i} a^2_{i,k}}.
\label{Dynamical content}
\end{equation}

\bibliography{refs}
\bibliographystyle{vancouver}